\documentclass[
    acmsmall,
    balance=true,
    nonacm=true,
    authorversion=true,
    urlbreakonhyphens=true
]{acmart}

\usepackage{amsmath}
\usepackage{amsfonts}
\usepackage{algorithmic}
\usepackage{graphicx}
\usepackage{textcomp}
\usepackage{xcolor}
\usepackage{acro}
\usepackage{pifont}
\usepackage{csquotes}
\usepackage{listings}
\usepackage{prettyref}
\usepackage[english]{babel}
\usepackage{caption}
\usepackage{subcaption}
\usepackage{xspace}
\usepackage{nicefrac}
\usepackage[inline]{enumitem}
\usepackage{upgreek}
\usepackage{textcomp}
\usepackage{wrapfig}

\DeclareCaptionLabelFormat{withfigure}{Fig. \arabic{figure}#2.}
\acsetup{
    single=true,
    first-style=long-short,
    subsequent-style=short,
}
\DeclareAcroEnding{possessive}{'s}{'s}

\NewAcroCommand\acg{m}{\acropossessive\UseAcroTemplate{first}{#1}}
\NewAcroCommand\acsg{m}{\acropossessive\UseAcroTemplate{short}{#1}}
\NewAcroCommand\aclg{m}{\acropossessive\UseAcroTemplate{long}{#1}}
\NewAcroCommand\acfg{m}{%
    \acrofull
    \acropossessive
    \UseAcroTemplate{first}{#1}%
}
\NewAcroCommand\iacsg{m}{%
    \acroindefinite
    \acropossessive
    \UseAcroTemplate{short}{#1}%
}
\DeclareAcronym{AI}{
    short=AI,
    long=artificial intelligence
}
\DeclareAcronym{AoT}{
    short=AoT,
    long=ahead-of-time
}
\DeclareAcronym{AGV}{
    short=AGV,
    long=autonomous guided vehicle
}
\DeclareAcronym{API}{
    short=API,
    long=application programming interface
}
\DeclareAcronym{ATP}{
    short=ATP,
    long=automatic train protection
}
\DeclareAcronym{CPU}{
    short=CPU,
    long=central processing unit
}
\DeclareAcronym{CS}{
    short=CS,
    long=chip select
}
\DeclareAcronym{DMA}{
    short=DMA,
    long=direct memory access
}
\DeclareAcronym{EXTI}{
    short=EXTI,
    long=external interrupt
}
\DeclareAcronym{FPGA}{
    short=FPGA,
    long=field-programmable gate array
}
\DeclareAcronym{GPIO}{
    short=GPIO,
    long=general purpose I/O
}
\DeclareAcronym{HAL}{
    short=HAL,
    long=hardware abstraction layer
}
\DeclareAcronym{IO}{
    short=I/O,
    long=input/output
}
\DeclareAcronym{IoT}{
    short=IoT,
    long=internet of things
}

\DeclareAcronym{IIoT}{
    short=IIoT,
    long=industrial IoT
}

\DeclareAcronym{ISR}{
    short=ISR,
    long=interrupt service routine
}
\DeclareAcronym{IRQ}{
    short=IRQ,
    long=interrupt request
}
\DeclareAcronym{IVT}{
    short=IVT,
    long=interrupt vector table
}
\DeclareAcronym{JIT}{
    short=JIT,
    long=just-in-time
}
\DeclareAcronym{JVM}{
    short=JVM,
    long=Java Virtual Machine
}
\DeclareAcronym{LL}{
    short=LL,
    long=low layer
}
\DeclareAcronym{MCU}{
    short=MCU,
    long=microcontroller unit
}
\DeclareAcronym{MISO}{
    short=POCI,
    alt=MISO,
    alt-acc=POCI,
    long=peripheral-out-controller-in
}
\DeclareAcronym{ML}{
    short=ML,
    long=machine learning
}
\DeclareAcronym{MMIO}{
    short=MMIO,
    long=memory mapped I/O
}
\DeclareAcronym{MMIO-C}{
    short=MMIO-C,
    long=MMIO Copy
}
\DeclareAcronym{MOSI}{
    short=PICO,
    alt=MOSI,
    alt-acc=PICO,
    long=peripheral-in-controller-out
}
\DeclareAcronym{PLC}{
    short=PLC,
    long=programmable logic controller
}
\DeclareAcronym{MPU}{
    short=MPU,
    long=memory protection unit
}
\DeclareAcronym{OPAL}{
    short=OPAL,
    long=OpenPOWER Abstraction Layer
}
\DeclareAcronym{OEM}{
    short=OEM,
    long=original equipment manufacturer
}
\DeclareAcronym{OS}{
    short=OS,
    long=operating system,
    short-plural=es
}
\DeclareAcronym{OSAPI}{
    short=OS-API,
    long=operating system's API
}
\DeclareAcronym{PCB}{
    short=PCB,
    long=printed circuit board
}
\DeclareAcronym{RAM}{
    short=RAM,
    long=random access memory
}
\DeclareAcronym{RAPI}{
    short=R-API,
    long=register API
}
\DeclareAcronym{RCU}{
    short=RCU,
    long=read-copy-update
}
\DeclareAcronym{RTOS}{
    short=RTOS,
    long=real rime operating system,
    short-plural=es
}
\DeclareAcronym{SOC}{
    short=SOC,
    long=system on a chip
}
\DeclareAcronym{SPI}{
    short=SPI,
    long=Serial Peripheral Interface
}
\DeclareAcronym{SRAM}{
    short=SRAM,
    long=static random access memory
}
\DeclareAcronym{TOCTOU}{
    short=TOCTOU,
    long=time-of-check to time-of-use
}
\DeclareAcronym{VM}{
    short=VM,
    long=virtual machine
}
\DeclareAcronym{WAMR}{
    short=WAMR,
    long=WebAssembly Micro Runtime
}
\DeclareAcronym{WASM}{
    short=Wasm,
    long=WebAssembly
}
\DeclareAcronym{WASI}{
    short=WASI,
    long=WebAssembly System Interface
}
\DeclareAcronym{WCET}{
    short=WCET,
    long=worst-case execution time
}
\DeclareAcronym{WOET}{
    short=WOET,
    long=worst-observed execution time
}

\newcounter{challenge}

\makeatletter

\newcommand{\unit}[2]{\ensuremath{#1\,\mathrm{#2}}}
\newcommand{\challenge}[1]{%
    \refstepcounter{challenge}
    \paragraph*{Challenge~\arabic{challenge} -- #1}%
}
\newcommand{\circledtext}[1]{%
    \raisebox{.5pt}{\textcircled{\raisebox{-.9pt}{#1}}}%
}
\newcommand{\fakesc}[2]{%
    #1\scalebox{.8}{#2}%
}
\newcommand{\resetacronmys}{%
    \acresetall%
    \skipcommonacronyms%
}
\newenvironment{fakewrapfig}{
\newlength{\parindent@original}
\setlength{\parindent@original}{\parindent}
\vspace{1mm}
\newcommand{\torightfromhere}{%
    \end{minipage}%
    \hfill%
    \renewcommand{\torightfromhere}{}{}%
    \begin{minipage}{.48\linewidth}
        \setlength{\parindent}{\parindent@original}
}%
\noindent%
\begin{minipage}{.48\linewidth}
    \setlength{\parindent}{\parindent@original}
}
{
\end{minipage}
}
\makeatother

\newcommand{\eg}{e.\,g.,\xspace}
\newcommand{\ie}{i.\,e.,\xspace}
\newcommand{\etal}{~et~al.\@\xspace}
\newcommand{\prologueepilogue}{prologue-epilogue\xspace}
\newcommand{\userspace}{user space\xspace}
\newcommand{\kernelspace}{kernel space\xspace}
\newcommand{\devicetree}{devicetree\xspace}
\newcommand{\devicetrees}{devicetrees\xspace}
\newcommand{\Devicetrees}{Devicetrees\xspace}
\newcommand{\ZAPI}{Zephyr \ac{API}\xspace}
\newcommand{\system}{\textsc{Wasm-IO}\xspace}
\newcommand{\systemfakesc}{\fakesc{W}{ASM}-IO\xspace}
\newcommand{\service}{{\ac{WASM} service}\xspace}
\newcommand{\services}{{\ac{WASM} services}\xspace}
\newcommand{\OEM}{\ac{OEM}\xspace}

\newcommand{\operators}{operators\xspace}

\newrefformat{chap}{Chapter~\ref{#1}}
\newrefformat{sec}{Section~\ref{#1}}
\newrefformat{fig}{Figure~\ref{#1}}
\newrefformat{tab}{Table~\ref{#1}}
\newrefformat{tbl}{Table~\ref{#1}}
\newrefformat{listing}{Listing~\ref{#1}}
\newrefformat{lst}{Listing~\ref{#1}}

\newlist{enumnone}{enumerate*}{1}
\setlist[enumnone]{label=}
\newlist{enumalpha}{enumerate*}{1}
\setlist[enumalpha]{label=(\alph*)}

\newcommand{\skipcommonacronyms}{\acuse{%
  API,%
  CPU,%
  FPGA,%
  GPIO,%
  IO,%
  JVM,%
  RAM,%
  RTOS,%
  SRAM,%
}}

\begin{abstract}
    \acswitchon

Containerization has become a ubiquitous tool in software development.
Due to its numerous benefits, including platform interoperability and secure execution of untrusted third-party code, this technology is a boon to industrial automation, promising to provide aid for their inherent challenges -- except one, which is interaction with physical devices.
Unfortunately, this presents a substantial barrier to widespread adoption.

In response to this challenge, we present \system, a framework designed to facilitate peripheral \ac{IO} operations within \ac{WASM} containers.
We elucidate fundamental methodologies and various implementations that enable the development of arbitrary device drivers in \ac{WASM}.
Thereby, we address the needs of the industrial automation sector, where a prolonged device lifetime combined with changing regulatory requirements and market pressure fundamentally contrasts vendors' responsibility concerns regarding post-deployment system modifications to incorporate new, isolated drivers.

In this paper, we detail synchronous \ac{IO} and methods for embedding platform-independent peripheral configurations within \ac{WASM} binaries.
We introduce an extended priority model that enables interrupt handling in \ac{WASM} while maintaining temporal isolation.
Our evaluation shows that our proposed \ac{WASM} isolation can significantly reduce latency and overhead.
The results of our driver case study corroborate this.
We conclude by discussing overarching system designs that leverage \system, including scheduling methodologies.

\end{abstract}

\setcopyright{acmlicensed}
\copyrightyear{2025}
\acmYear{2025}
\acmDOI{XXXXXXX.XXXXXXX}

\acmJournal{TECS}
\acmVolume{0}
\acmNumber{0}
\acmArticle{0}
\acmMonth{0}

\begin{document}

\skipcommonacronyms

\title{Extending Lifetime of Embedded Systems by WebAssembly-based Functional Extensions Including Drivers}

\author{Maximilian Seidler}
\authornote{Also affilliated with Siemens AG.}
\affiliation{%
  \institution{Friedrich-Alexander-Universität Erlangen-Nürnberg}
  \city{Erlangen}
  \country{Germany}
}
\email{maximilian.seidler@fau.de}
\orcid{0009-0007-1601-9311}

\author{Alexander Krause}
\affiliation{%
  \institution{Technische Universität Dortmund}
  \city{Dortmund}
  \country{Germany}
}
\email{alexander.krause@tu-dortmund.de}

\author{Peter Ulbrich}
\affiliation{%
  \institution{Technische Universität Dortmund}
  \city{Dortmund}
  \country{Germany}
}
\email{peter.ulbrich@tu-dortmund.de}



\received{20 February 2007}
\received[revised]{12 March 2009}
\received[accepted]{5 June 2009}

\maketitle

\resetacronmys

\section{Introduction}\label{sec:introduction}

Containerization has become increasingly valuable, as it provides a range of advantages that extend from simplified and cost-sensitive deployment and scalability to enhanced security.
Developers can construct, test, and deploy applications from various sources capable of operating reliably across different devices by encapsulating applications within containers, provided a suitable runtime framework is available.

These advantages are particularly compelling in the context of \ac{IIoT} and industrial automation, where \acp{OEM} are driven to deliver robust, scalable, and secure execution platforms, commonly referred to as \acp{PLC}.
These platforms provide a basic framework for operators to develop their applications.
The operational area of \acp{PLC} is diverse, encompassing any process involving control ranging from plant managing via \ac{AGV} steering to facility management.
As a result, they span minimal to substantial computational capacities and rely on sensors, actuators, and communication.
However, this work aims towards application in the process industry.

\acp{PLC} play a critical role in the process industry, \eg in regulating the speed of conveyor belts, actuating diverters or valves, and controlling mixers.
Sensors provide inputs, \ie fill levels, temperatures, or pressure of fluids of vessels.
Production processes are performed on batches or continuous material characterized by generally slow operational speeds (hundreds of milliseconds to minutes).
However, risks are significant in emergencies.

The production of margarine serves as an illustrative example.
Oil is extracted from oilseeds, followed by purification, deacidification, and blending with various ingredients.
The oil is thickened through hydrogenation, pasteurized, and finally packaged for distribution. The entire production process operates on a continuous material basis and typically features slow operational speeds, ranging from hundreds of milliseconds to several minutes.
Nonetheless, the system must be capable of addressing emergency situations in a timely manner.

Accidentally mixing acid and alkaline additives can engage in exothermic reactions and precipitation, contaminating the pipes and batches produced.
Moreover, leaks can result in the escape of high-pressure and hot fluids or gases, including hydrogen, which is utilized in the thickening phase.
Pyrophoric catalysts and the risk of spontaneous ignition for certain oils impede rapid emergency shutdowns and sealing measures.
Additionally, in other production environments, byproducts may even be toxic, while solidification or crystallization processes can risk demolishing the entire plant.

This example illustrates the rationale for operators designing these systems to remain largely unchanged for extended periods -- often for years or even decades.
Although production parameters may undergo minor adjustments, the fundamental nature of the process remains stable, in contrast to manufacturing assembly lines.
Additionally, maintaining the functional safety of emergency routines within the plant is essential.

However, competitive pressures encourage modifying these systems.
Implementing precise regulation of temperature gradients during mixing can enhance emulsion stability.
Moreover, extensive data collection throughout the storage and preprocessing phases facilitates fine-tuning parameters, ultimately improving product quality and reducing the need for quality control measures or refinements, lowering overall production costs.
Predictive maintenance of moving components similarly contributes to operational efficiency in manufacturing processes.
Moreover, evolving regulatory landscapes may impose new requirements on these systems.
For instance, changing health regulations mandate extended protocolling of production conditions, while occupational safety regulations may demand surveillance or enhanced plant monitoring.

Although these factors significantly encourage the adoption of new software and sensors, \ie peripherals, they simultaneously introduce organizational and practical complexities, particularly regarding reliability and available information.
Undertaking firmware modifications or reflashing new firmware undermines guarantees provided by the initial software certification process.
Coupled with considerable deployment efforts and the inherent risk of disrupting ongoing production -- which may lead to significant financial loss -- \acp{OEM} often prohibit operators from system modifications within the bounds of their liability.
Furthermore, the absence of comprehensive documentation, such as detailed hardware configurations or clearly defined scheduling algorithms, worsens these challenges.

This is where the flexibility and modularity of containerization become invaluable.
New software functionalities and hardware drivers, which must be considered harmful from the \acg{OEM} perspective, may be retrofitted, leveraging isolated containers.
Containers enable operators to integrate software without modifying existing, proven and certified components -- contingent upon the firmware being designed to provide suitable mechanisms from the outset.
However, in contrast to general-purpose computing environments, the rigorous real-time requirements for emergency cases imposed by the process industry must remain uncompromised.

\ac{WASM} has recently emerged as a potential alternative or complementary technology for containerized applications, including those in safety-critical embedded systems.
The compact, binary-compiled format and efficient execution model of \ac{WASM} enable its operation on resource-constrained devices, such as \acp{PLC}, while maintaining the isolation benefits inherent to container technology.
Its capacity to execute in a secure sandboxed environment renders it particularly well-suited to industrial automation, where security and stability are paramount.
Interaction with the host system is restricted, ensuring program faults do not propagate and preventing exceptional outages.
Redeploying is simplified to loading new binaries without the necessity of reflashing or recompiling the firmware or restarting the \ac{PLC}.
\acg{WASM} platform-agnostic design enables it to run across diverse hardware and software environments, facilitating its portability -- an indispensable attribute for \acp{OEM} overseeing expansive, heterogeneous deployments and long life cycles.
As \acl{WASM} utilization proliferates, it is becoming a promising technology that addresses some of the constraints of conventional containers in industrial contexts, enabling \acp{OEM} to develop scalable, efficient, and secure software solutions for automation applications~\cite{liuAerogelLightweightAccess2021,peachEWASMPracticalSoftware2020,wallentowitzPotentialWebAssemblyEmbedded2022,liThingSpireOSWebAssemblybased2021,singhWARDuinoDynamicWebAssembly2019,wenWasmachineBringIoT2020,makitaloWebAssemblyModulesLightweight2021,liWiProgWebAssemblybasedApproach2021,peachEWASMPracticalSoftware2020}.
We elaborate on the specific properties of \ac{WASM} in \prettyref{sec:background}.

\subsection{Problem Statement}

While \ac{WASM} offers the abovementioned advantages, its adoption in safety-critical and real-time environments is still early.
In particular, it is limited in \ac{IIoT} settings by the absence of direct support for low-level peripheral device interactions and interfaces due to the security and storage model of \ac{WASM}.
\acg{WASM} isolation mechanisms impede \ac{IO}, as drivers necessitate access to device registers for \ac{MMIO} operations that are inherently memory-unsafe.
Even in type-safe programming environments, register addresses are ultimately derived from interpreting numeric register addresses as referenceable types.
Moreover, device-specific peripheral registers fundamentally conflict with the portability principles of \ac{WASM}.
However, enabling drivers in \ac{WASM} is crucial for updating and managing long-lifecycle field devices, particularly when new hardware extensions are involved.

In designing future-proof \ac{WASM}-based automation systems that facilitate the post hoc integration of drivers, it is essential to prioritize real-time consistency.
This necessitates an upfront consideration of the temporal impacts associated with drivers introduced post-deployment.
However, drivers frequently employ interrupts to manage asynchronous events, significantly impacting the overall considerations of the system's determinism.
At present, \ac{WASM} lacks general support for interrupt handling.
Further, no runtime environment provides an execution model that ensures deterministic system behavior.
These observations underscore the necessity for mechanisms that facilitate \ac{WASM} in real-time \ac{IIoT} and automation applications, which demand low interrupt latencies and predictable overheads.

Consequently, this paper investigates utilizing \ac{WASM} in real-time industrial systems in a secure, efficient, and temporally well-behaved manner.
It assesses solutions for \ac{IO} operations and interrupt handling within \ac{WASM}, along with strategies for deterministic control of interrupt latencies.
The primary objective is to provide a framework for integrating modifications, including drivers for peripherals added post-deployment.
To address these complexities, we have identified several key challenges.

\challenge{Platform-agnostic Device Addressing}

Granting \ac{WASM} executables access to generic \ac{IO} requires expressing peripheral requirements in a platform-agnostic manner.
This principle extends to the use of platform-specific \ac{IO} registers utilized for control operations.

\emph{Our Approach:}
We propose to embed \ac{IO} requirements directly into the \ac{WASM} binary as immutable name-based references to registers.
At load time, our novel mechanism enables the matching of these requirements with the host's actual \ac{IO} capabilities.

\challenge{Utilization of \ac{IO}}

\ac{WASM} lacks native instructions for peripheral interaction and prevents manipulation of register addresses due to its isolation model.
Consequently, alternative mechanisms for peripheral interaction must be provided while ensuring that the host system retains complete control over secure access to these resources at all times.

\emph{Our Approach:}
We present two methods for interaction.
First, we offer an \ac{API}, and second, the \ac{WASM} runtime has been modified to enable utilization of \ac{IO} registers within the container's address space and intercept on access.
Both methods check and redirect accesses to actual hardware registers based on permissions.

\challenge{Controlling System Determinism}

Handling interrupts in real-time systems necessitates careful management of priority levels to maintain predictability.
Latency-sensitive tasks, particularly those related to hardware, require elevated priority, especially in the context of drivers.
However, the unpredictable timing behavior of \ac{WASM} drivers introduces two problematic scenarios: limiting interrupt handling to the application level compromises timeliness while permitting interference with other \ac{OS}-level drivers undermines deterministic behavior.

\emph{Our Approach:}
We introduce an extended priority model incorporating a conventional prologue and epilogue found in contemporary \acp{OS} and a specialized \ac{WASM}-specific epilogue level.
This structure enables the sequentialization of \ac{WASM} handlers while preserving deterministic latencies for firmware interrupts and temporal isolation.

\challenge{Costs and Predictability}

To assess the practicality of our solution, it is essential to quantify the overheads associated with synchronous \ac{IO} and understand the temporal impact of interrupts on the system.
Our primary challenge is to determine overhead, worst-case execution costs, and establish upper bounds for execution paths.

\emph{Our Approach:}
We developed multiple \ac{IO} access variants, executed at the kernel level and in user space and conducted measurement-based timing analysis to characterize them.
We also identified worst-case paths, which, in conjunction with a hardware cost model, allow estimating \acp{WCET} in varying scenarios.

\subsection{Contributions and Outline}

With \system, we introduce a new approach for retrofitting drivers and interrupt handlers into the \ac{WASM} ecosystem.
Our work makes four significant contributions:

\begin{enumerate}
	\item \textit{Device Configuration and \ac{IO} Support:} \system offers
    \begin{enumalpha}
		\item a platform-agnostic device configuration concept
		\item secure memory access with multiple variants to support trusted \ac{IO} in \ac{WASM}.
	\end{enumalpha}
	\item \textit{Interrupt and Execution Model:} \system extends traditional priority and execution models to enable systematic and predictable interrupt handling within \ac{WASM} containers.
	\item \textit{Cost and Latency Analysis:} We provide a thorough, measurement-based analysis of overheads, \acp{WOET}, latencies, and worst-case paths for each implementation variant, establishing a reliable foundation for utilizing \ac{WASM} in \ac{IIoT} and real-time applications.
	\item \textit{Prototype in Zephyr \acs{RTOS}:} \system's prototype in Zephyr \acs{RTOS} includes optimized, analyzable mechanisms for memory access and interrupt handling, minimizing runtime checks and overhead.
\end{enumerate}

The remainder is organized as follows:
We present relevant background in \prettyref{sec:background}.
The concepts are detailed in \prettyref{sec:concept}, and their implementation is presented in \prettyref{sec:implementation}.
\prettyref{sec:eval} provides our comprehensive evaluation on \system's implementation variants.
We discuss \system's envisioned usage and its limitations in \prettyref{sec:discussion}.
Related work is addressed in \prettyref{sec:related_work}.
\prettyref{sec:conclusion} concludes our work.

\section{Background}\label{sec:background}

This section provides background information on \ac{OS} specifics, \devicetrees, and \ac{WASM} and its isolation concepts.

\subsection{Traditional System Model}\label{sec:interrupt_handling_in_os}

\begin{wrapfigure}{L}{0.5\textwidth}
	\vspace{-.2cm}
	\includegraphics[width=\linewidth]{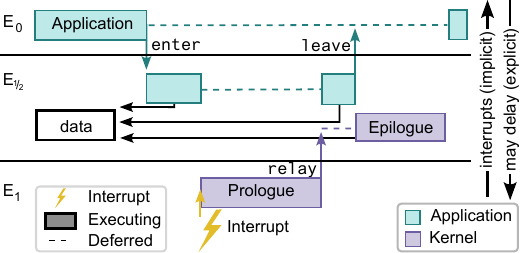}
	\caption{An example pro-epilogue interrupt synchronization according to the priority level model.}
	\label{fig:proepilogue}
	\vspace{-1cm}
\end{wrapfigure}

\ac{OS}-level interrupt handling and synchronization follow a strict priority model~\cite{schroder-preikschatLogicalDesignParallel1994} with three priority levels: E$_i$ $i \in \{0,\nicefrac12,1\}$.
A level E$_i$ obeys the following rules:
\begin{enumerate}
\item It can be interrupted at any time from E$_j$ for $j > i$.
\item It is not interrupted by control flow on E$_k$ for $k \leq i$.
\item Control flows on E$_i$ run sequentially (\textit{run to completion}).
\end{enumerate}

As illustrated in \prettyref{fig:proepilogue}, the level E$_0$ is utilized for the application code.
In contrast, any \ac{ISR} runs on E$_1$, and thus can interrupt levels E$_0$ and E$_{\nicefrac12}$.

To mitigate the impact of system unresponsiveness due to interrupts being disabled during ISR execution, a two-stage approach is employed.
A prologue on E$_1$ is dedicated to time-critical operations such as reading data from devices.
It may issue an epilogue on E$_{\nicefrac12}$, which handles data processing.
The relocation of the epilogue to E$_{\nicefrac12}$ permits the re-enabling of interrupts.
Epilogues are sequentialized, and pending epilogues are placed in a queue by the OS, ensuring completion before returning control to the application level (E$_0$) as shown in \prettyref{fig:proepilogue}.
The Linux implementation refers to the prologue as the \emph{top half} and the epilogue as the \emph{bottom half}~\cite{love:2010}.

Data to be shared between handlers and application tasks logically resides at epilogue level E$_{\nicefrac12}$.
Due to sequentialization rules, no additional synchronization is needed when executed on a common core~\cite{schroder-preikschatLogicalDesignParallel1994}.

\subsection{\Devicetrees}

The \textit{\devicetree} specification~\cite{DevicetreeSpecificationV04rc12021} addresses the need for a common understanding of hardware interfaces in computer systems, particularly embedded devices.
It defines a hierarchical tree structure where each node represents a device or component and contains properties describing its characteristics, including register addresses and configuration.
\Devicetrees can also represent interrupt sources, controllers, and numbers as node properties.
Nodes can be identified by their \emph{labels} or hierarchical paths.
While not explicitly stated, the hierarchical structure naturally aligns with hardware levels: CPU, \ac{SOC}, and peripherals.
The first two are considered immutable due to requiring board modifications, while the peripheral level allows for flexible replacement of sensors, actuators, and other external devices.
In Zephyr, which is the basis for our prototype, the \devicetree serves exclusively for software design and compilation during the prototyping phase.
Notably, there is no corresponding runtime equivalent on the hardware board~\cite{DevicetreeZephyrProject}.

\subsection{\acl{WASM} Fundamentals}

\acf{WASM}~\cite{haasBringingWebSpeed2017,rossbergWebAssemblySpecification102022} was introduced in 2017 by major browser vendors as a performance-focused alternative to JavaScript.
Designed as a low-level, formally defined binary instruction format, \ac{WASM} emphasizes fast, portable execution with critical features like memory isolation and hardware independence.
LLVM-based toolchains facilitate the generation of binaries from multiple programming languages, including C/C++, Rust, and others.
This flexibility has led to \acg{WASM} adoption in fields beyond web development, including embedded and edge computing.

\acl{WASM} bytecode can be compiled \ac{AoT} or \ac{JIT} to native instructions or directly executed via interpretation.
In all cases, a dedicated runtime is utilized.
Since \ac{CPU} instruction set is abstracted by the runtime, \ac{WASM} bytecode is inherently portable across diverse hardware platforms.
However, this indirection introduces execution time overhead, which varies based on runtime optimization (such as \ac{JIT} or \ac{AoT} compilation) and the specific implementation on embedded devices.
In the literature~\cite{wallentowitzPotentialWebAssemblyEmbedded2022,liuAerogelLightweightAccess2021,menetreyTwineEmbeddedTrusted2021,wangHowFarWeve2022,wenWasmachineBringIoT2020,peachEWASMPracticalSoftware2020,zaeskeWebAssemblyAvionicsDecoupling2023}, the overhead factors for microcontrollers range from $1.14$ to $36.15$ in comparison to native performance.
The lower end of this range is achieved only through \ac{AoT} compiling, while interpreters typically exhibit an overhead factor greater than ten.
More powerful edge devices demonstrate similar overheads.

As a stack-based virtual machine, \ac{WASM} instructions manipulate an operand stack, allowing values to be popped, manipulated, and pushed.
Per the \ac{WASM}~1.0 specification~\cite{rossbergWebAssemblySpecification102022} and its \ac{WAMR} implementation~\cite{WAMRMemoryModel2023,UnderstandWAMRStacks2023}, a \ac{WASM} executable consists of a \emph{module} defining static components, such as bytecode, function definitions, tables, and global values.
When instantiated, a \emph{module instance} and an \emph{execution environment} are created with initialized symbols and memory (\textit{the linear memory}).
Execution environments then maintain a call stack, operand stack, and references to both the module instance and the current function.

\ac{WASM} provides mechanisms to interact with the host system by \emph{importing} functions.
Imported functions are used like internal functions but transfer control from \ac{WASM} bytecode to native code.
Ensuring memory isolation remains the responsibility of the host system.
The \ac{WASI}~\cite{gohmanWebAssemblyWASIV0222024} has emerged as a standard for defining system interfaces, offering file system access, and other low-level interactions for non-web environments.
\ac{WASI} and other evolving interface specifications aim to improve \acg{WASM} portability and usability for embedded applications.

\subsection{Isolation in \acl{WASM}}\label{sec:isolation_in_wasm}

\acl{WASM} provides robust spatial isolation through a combination of its memory model and type system.
\ac{WASM} bytecode can only access its own dedicated linear memory, \ie its heap, through \unit{32}{bit}
\footnote{
\unit{64}{bit} memory indices are subject of an official \ac{WASM} standard modification proposal~\cite{WebAssemblyProposalMemory642024} in phase four, approaching formal standardization.
} integer indices.
The type system prevents out-of-bounds accesses at the lower limit, while runtimes ensure the upper memory limit.
Accessing the host system's memory is not feasible.
Pointers are represented as \unit{32}{bit} offsets into global tables, such as the function table.
This eliminates dereferenceable types and, thus, the risk of malicious memory access.

Type consistency is enforced through static checks.
Control flow is protected by keeping the call stack separate from the operand stack, with control instructions constrained to prevent stack-based attacks~\cite{haasBringingWebSpeed2017}.

\acg{WASM} formal specification supports the mechanization and, thereby, the verification of bytecode and type system correctness~\cite{wattTwoMechanisationsWebAssembly2021,wattMechanisingVerifyingWebAssembly2018} and facilitates the development of verified runtime interpreters~\cite{wattWasmRefIsabelleVerifiedMonadic2023}.
However, verification remains an open question for \ac{AoT} and \ac{JIT} compiling~\cite{zaeskeWebAssemblyAvionicsDecoupling2023}.
Though current (industrial) implementations prioritize performance over formal correctness, we expect future runtimes to close this gap.
This paper assumes, where relevant, a correct runtime.

\section{\system: Design \& Concept}\label{sec:concept}

From the outset, it is evident that the scenario presented possesses several unique requirements.
The objective is to facilitate post-deployment modifications to \acp{PLC} to incorporate new, isolated drivers.
From a liability standpoint, \acp{OEM} must prohibit operators from modifying the firmware or any trusted component and ensure that operators cannot execute any untrusted code without proper memory isolation.
Moreover, faults must not propagate and entail exceptional stops.
Given that operational periods may extend over decades, reducing the maintenance associated with compiler toolchains is beneficial, and it may be achieved by utilizing one single standardized bytecode capable of running on all \acp{PLC} platform independently.
Furthermore, potentially incomplete documentation related to system setup, \ie scheduling configurations, encourages timely isolation.
The following section provides an in-depth analysis of the concepts and design principles underlying \system, the framework developed to meet the outlined requirements.

\subsection{Architecture}\label{sec:architecture}

\begin{wrapfigure}{R}{.5\textwidth}
  \includegraphics[width=\linewidth]{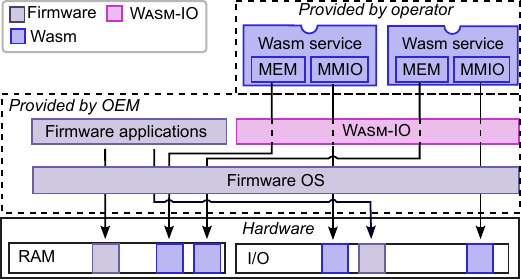}
  \caption{
      \system's core architecture with \services for third-party untrusted code that is loadable after device shipping.
  }
  \label{fig:concept}
\end{wrapfigure}

In our work we assume an embedded device, \ie \ac{PLC}, being assembled by an \OEM.
For the device to function, the \OEM provides a firmware, as depicted in \prettyref{fig:concept}.
This encompasses an \ac{OS} providing threading capabilities, a scheduler, device drivers, and other essential functionalities.
Firmware applications represent the initial business logic that the \ac{PLC} is shipped with.
Each software component is delivered with timing bounds and is subject to static verification and certification.

To extend the \acg{PLC} capabilities, \operators may load services that are not part of the firmware.
Therefore, it is necessary to consider the service's code as untrusted.
We introduce \system as an intermediate layer on kernel level for execution of untrusted services.
We utilize \ac{WASM} as the binary format, leveraging \acg{WASM} inherent sandboxing properties.
\acp{OEM} provide the \ac{WASM} runtime integrated into \system.
As the runtime is part of the trusted computing base, we assume formal correctness (see \prettyref{sec:isolation_in_wasm}).
Therefore, it is allowed to execute services in \kernelspace without the risk of system interference.

As the aforementioned services are provided as \ac{WASM} binaries, they are referred to as \emph{\services} in the following.
Each \service encompasses business logic, driver logic, or both.
It is single-threaded and isolated from other \services and the \ac{OS}.
An \ac{OS} thread thread executes the \ac{WASM} runtime, which in turn executes the \ac{WASM} binary.
While the runtime provides temporary, isolated \ac{RAM} parts for the \service to operate, \system exclusively assigns \ac{MMIO} regions to each service, if required.
It is important to note that assigning devices and \services exclusively is not a restriction.
\services are designed to support peripherals that were not present at firmware design.
Consequently, no conflicts with the firmware can occur.

When compiling a source program to \ac{WASM}, the control flow is preserved.
Notably, structures, \eg driver registers, are maintained if the address value is unknown to the compiler, preventing constant propagation.
We achieve this through our method of expressing peripheral requirements.
Additionally, there are some subtle differences in semantics.
In particular, \ac{WASM} lacks a mechanism to represent volatile variables and does not provide memory consistency principles, concepts that are commonly present in C/C++-based drivers.
However, compiler optimization is performed on an intermediate representation, typically LLVMIR, \emph{before} converting to \ac{WASM}, where these concepts are still present.
Since interpreters directly utilize this generated \ac{WASM}, memory access order is preserved.
For subsequent compilation processes, such as \ac{JIT} and \ac{AoT} compilation, the volatile behavior can be emulated through the use of atomics~\cite{heejinahnWebAssemblyToolconventionsVolatile2028}, which have yet to be standardized
\footnote{
Atomics is a proposed change to the \ac{WASM} specification, arising from the threads proposal~\cite{wattWebAssemblyProposalThreads2024} in phase four.
}.

\begin{wrapfigure}{r}{.5\textwidth}
  \includegraphics[width=\linewidth]{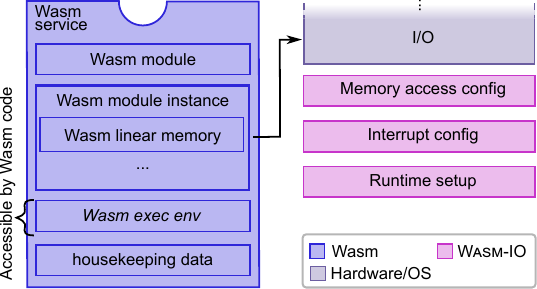}
  \caption{
      \system's core architecture from the memory perspective.
  }
  \label{fig:memory-setup}
\end{wrapfigure}

The relevant parts of the layout of \system are depicted in \prettyref{fig:memory-setup}.
The \emph{runtime setup} is utilized by \system to manage the initialization process of the \ac{WASM} runtime.
Each service consists of
\begin{enumnone}
    \item the \textit{\ac{WASM} module}, \ie the binary,
    \item the \textit{\ac{WASM} module instance}, \ie global storage
    \item the \textit{\ac{WASM} execution environment}, \ie the state of the stack machine, and
    \item some housekeeping data.
\end{enumnone}
The \emph{memory access configuration} includes mapping registers to a \service to assign \ac{MMIO} regions.
The following section elucidates details on this access configuration and generic \ac{IO}.
The \emph{interrupt config} enables asynchronous interaction, \ie interrupt handling, which is covered subsequently.
The operator's untrusted code, provided as \ac{WASM} binaries, is restricted to accessing only those memory regions within the execution environment as described in \prettyref{sec:isolation_in_wasm}.
In particular, the \services are unable to modify the peripheral configurations.

\subsection{Peripheral Identification and Access}

The ability to access \emph{arbitrary} devices from \emph{arbitrary} \services is contingent upon the resolution of several subproblems:
\begin{enumalpha}
  \item ensuring the availability of peripherals required by the \service on the device,
  \item embedding them in a hardware-agnostic manner, and finally
  \item enabling usage within the \service in the driver's assumed layout.
\end{enumalpha}

\paragraph{a) Ensuring Availability}\label{sec:ensure-existence}
\system enforces explicit permission from the \OEM at design time to use a peripheral post hoc within a \service, which enforces the separation of firmware drivers and \service drivers.
Thereby, responsibility is assigned to the \OEM to ensure future compatibility and freedom of interference.
Furthermore, it assures the exclusive assigning of peripherals at all times, as assumed in \prettyref{sec:architecture}.
Accordingly, peripheral registers (or devices) are attached with a label and, together with the physical address (or \ac{OS} driver instance), are written in the memory access configuration (see \prettyref{fig:memory-setup}).
We denote these as \emph{exposed registers} (or \emph{devices}).
Our implementation (\prettyref{sec:implementation}) shows how this naturally aligns with the concept of \devicetrees if our novel \devicetree binding is introduced.
Upon loading a \service, \system will verify that all registers (or devices) requested by the service are exposed.
If any of the requested dependencies is unavailable, the \service will be rejected, thus preventing any potential runtime errors.

\paragraph{b), c) Expressing Peripheral Requirements and Usage}
A driver for a novel peripheral device is typically programmed against a \ac{HAL}.
In order to facilitate board-independent development of driver logic, the \ac{HAL} provides placeholders for each register address.
The placeholders are resolved at compile time employing a board-specific configuration.
Consequently, the resulting binary only contains physical addresses.
\system builds upon this established approach.
Rather than resolving placeholders to the physical addresses, \system introduces dummy registers for each placeholder linked to a unique register label.
These dummy registers remain within the retained \ac{WASM} binary when compiling a \service.
Upon loading the \service, the each dummy register is resolved to a physical address using the memory access configuration.
\system offers a build tool that streamlines this process.

\subsection{Interrupt Handling}
\begin{figure}[t]
    \includegraphics[width=\linewidth]{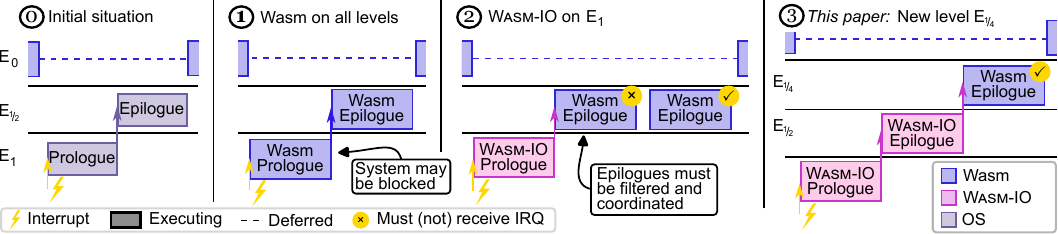}
    \caption{
        The deduction of the new extended \prologueepilogue model, with step~\circledtext{1} blocking system responses and step~\circledtext{2} lacking a trusted entity to ensure proper filtering of the \ac{WASM} epilogues.
        Step~\circledtext{3} presents the final solution.
    }
    \label{fig:prologue-epilogue-wasm}
  \end{figure}

Communication from a specific device to its corresponding driver occurs asynchronously via interrupts.
As previously outlined in \prettyref{sec:interrupt_handling_in_os}, interrupt handling in contemporary \acp{OS} is divided into two phases: the prologue facilitates communication to the device, i.e. data reception, and the epilogue performs actual data processing.

This initial situation is illustrated in \prettyref{fig:prologue-epilogue-wasm} as step~\circledtext{0}.
The \service operates on level E$_0$, allowing it to be deferred, while any prologue or epilogue is executed on level E$_1$ or E$_{\nicefrac{1}{2}}$, respectively.

Step~\circledtext{1} illustrates the \prologueepilogue model being applied naively to the system.
As a \service executes arbitrary code, it is evident that allowing it to run its own prologue and epilogue introduces significant isolation vulnerabilities.
Untrusted code is executed in the prologue with the highest priority, and interrupts being disabled.
Consequently, any program failure inevitably results in a non-recoverable device failure, \eg a division by zero occurring with disabled interrupts necessitates external hardware for reset.
Moreover, system determinism is invalidated due to the uncertain execution time and termination.
As interrupts are disabled, they can only be enforced for inherently slow interpreters.
Therefore, \system offers a general-purpose \emph{\systemfakesc prologue} that retrieves data from the device on behalf of the \service.
To facilitate platform-independent interrupts, the same concepts apply as for the exposed registers.
\acp{OEM} define the method for clearing interrupt flags during the design phase.
\services can register to receive \acp{IRQ} by specifying an interrupt label along with a priority level.
This priority facilitates the coordination of multiple \ac{WASM} epilogues that may arise from interrupt sharing.

As shown in step~\circledtext{2}, this eliminates the vulnerability inherent in the prologue.
However, multiple peripherals may share a single interrupt line, leading to the execution of all \ac{WASM} epilogues associated with that line, regardless of their permissions to the respective peripherals.
Additionally, their proclaimed priority is employed to enforce consistent sequentialization in scenarios with multiple \ac{WASM} epilogues.
Furthermore, the execution of firmware epilogues must take precedence over \ac{WASM} epilogues.
Thus, incorporating an additional trusted entity is essential to ensure effective filtering and proper ordering.

To address this issue, we introduce a common \emph{\systemfakesc epilogue} and shift the \ac{WASM} epilogues to a novel priority level E$_{\nicefrac{1}{4}}$, situated between the \system epilogue and the \service.
This change results in \system's \prologueepilogue model depicted in step~\circledtext{3}.
The \system epilogues may supersede the \ac{WASM} epilogues, allowing for control over the \ac{WASM} epilogues.
Firmware epilogues continue to reside on E$_{\nicefrac{1}{2}}$.

To provide data reception on the prologue level, we facilitate the \service to register data to be received alongside the corresponding interrupt.
If accessible to the \service, this data is buffered.
Prior to scheduling a specific service's epilogue, the \system epilogue copies data to the \ac{WASM} linear memory with the dummy registers modified accordingly.
As the \service then operates on this \emph{copy}, the necessity for the \ac{WASM} epilogue to enter any level below E$_{\nicefrac{1}{4}}$ to retrieve the data is eliminated.
The primary objective of this method is to significantly reduce interrupt latencies for the \services, particularly with \userspace enabled.
Programmatic data reception must be performed in the \ac{WASM} epilogue.

Additionally, with our revised data model, this approach facilitates timely and deterministic interrupt synchronization.
A further advantage of restricting a \service to level E$_{0}$ and E$_{\nicefrac{1}{4}}$ is the avoidance of explicit interrupt synchronization, as described in \prettyref{sec:interrupt_handling_in_os}.
\system assumes responsibility for \begin{enumnone}
    \item data reading (\system prologue),
    \item movement across the different priority levels, and
    \item availability in the \ac{WASM} epilogue.
\end{enumnone}

\section{Implementation}\label{sec:implementation}

We built our prototype of \system based on Zephyr \ac{RTOS} and the classic \ac{WAMR} interpreter as interpreters can generally be verified, and no such compilers are on the horizon.
Consequently, despite its adverse impact on performance, we have pursued the interpreter-based approach.
Additionally, due to limitations in \ac{RAM}, we adjusted the page size of \ac{WASM}.

\subsection{Peripheral identification}\label{sec:peripheral-identification}

We utilize \devicetrees as a well-known and suitable data structure to create the memory access configuration.
We introduce a new \devicetree binding for this purpose.
During firmware design, \acp{OEM} must identify all nodes, such as unused buses or \acp{GPIO}, where peripherals may be connected after deployment.
These nodes are referenced explicitly in a node of the new binding type.
Our build tool then generates unique, immutable names derived from the nodes' immutable path or label, taking advantage of the hierarchical structure.
Along with selected properties, \ie interrupt numbers and register addresses, these names comprise the build-time part of the memory access configuration.
We recommend using node labels, such as \enquote{spi1}, instead of full paths to improve cross-platform compatibility.

When adding new peripherals and drivers post-deployment, they can identify the labels of the registers or interrupts they depend on by following a similar procedure.
These new components can virtually be appended to the immutable parent elements in the \devicetree.
Our build tool then consistently generates dependencies with the same labels.
These labels are then embedded into the custom section of the \ac{WASM} binary.
At load time, the dependencies in the \ac{WASM} binary can be matched with the build-time components of the memory access configuration to provide the physical addresses or interrupt numbers.

Note that \devicetrees are a convenient data structure for consistently generating register or device labels.
However, our approach is not strictly reliant on it.
Consequently, we argue that porting our approach to different \acp{RTOS} requires minimal effort.

\subsection{Synchronous Peripheral Interaction}

\begin{figure}[t]
	\centering
    \includegraphics[width=\textwidth]{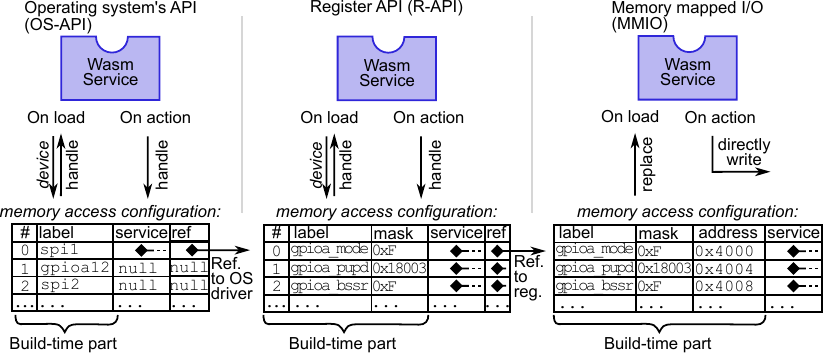}
    \caption{Schematic of different peripheral access types forming the implementation variants in \system.}
    \label{fig:approaches}
\end{figure}

In the absence of support for peripheral interaction in \ac{WASM}, \system provides one reference and four novel methods for synchronous peripheral access:
\acs{OSAPI}, \acs{RAPI}, \acs{MMIO}, and \acs{RAPI} and \acs{MMIO} with \ac{DMA} acceleration.
These will be detailed in the following.

\paragraph{\Acf{OSAPI}}
The \ac{OSAPI} is based on existing literature~\cite{singhWARDuinoDynamicWebAssembly2019,liBringingWebassemblyResourceconstrained2022,kurteEmbeddedWASM2023} and serves as a reference baseline for enabling a \service to access an \ac{OS} peripheral driver \ac{API}.
To use this \ac{API}, a \service must be programmed against the existing \ac{OS} driver \ac{API}.

Device dependencies embedded in the \ac{WASM} binary refer to the \devicetree node rather than its registers, ensuring the existence of the device (see \prettyref{sec:ensure-existence}).
At load time, the \service requests a device handle corresponding to the desired driver instance instead of resolving dummy registers.
This method is illustrated on the left of \prettyref{fig:approaches}.
If the request is approved, the handle is stored in the memory access configuration (refer to \prettyref{fig:memory-setup}).

During runtime, the \service calls the \ac{OS}-internal \ac{API} for each \ac{IO} action, \ie \lstinline[breaklines=false, breakatwhitespace=false]!spi_write(...)!.
\system checks whether the service has the necessary privileges to the handle by reviewing the memory access configuration.
If access is granted, the request is forwarded to the corresponding driver.

Since this approach relies entirely on the \ac{OS}-specific driver and its configuration, the \ac{OSAPI} does not permit arbitrary peripheral access, which is the primary objective of this work.
Therefore, it is intended solely as a reference.

\paragraph{\Acf{RAPI}}
To facilitate access to \emph{arbitrary} peripherals, we introduce the novel \ac{RAPI}.
This approach refines the presented concept of an \ac{API}, allowing for direct manipulation of device registers.
Unlike the \ac{OSAPI}, where each device handle corresponds to an \ac{OS} driver, in \ac{RAPI}, each handle directly corresponds to a specific register.
As a result, a service can interact with any device as long as the corresponding register addresses are known and exposed.

The addresses in question are determined using dummy register labels, as \prettyref{sec:peripheral-identification} explains.
Consequently, interacting with a register involves calling a function instead of dereferencing a pointer, like in C-like languages.
However, this design enables \system to verify every register access by consolidating the memory access configuration.
In addition, this allows programming languages without a reference concept to be utilized.
The process is illustrated in the center of \prettyref{fig:approaches}.

For precise control, \system applies masks to the values prior to access, \eg when manipulating \ac{GPIO} registers, as they provide access to multiple \ac{IO} pins.
These masks are obtained from the initial \devicetree during the firmware build process and stored alongside the register addresses in the memory access configuration.
Registers are sorted by their usage frequency.
In one of our specific evaluation case where \system operates within \userspace (untrusted scenario), the \ac{API} call is equivalent to a system call.

\paragraph{\Acf{MMIO}}
As demonstrated in our evaluation (see \prettyref{sec:eval_synchronous_peripheral_interaction} and \prettyref{sec:case-study-spi}), both the \ac{OSAPI} and the \ac{RAPI} exhibit considerable latency.
We designed a significantly faster variant to address this issue by enabling true \acl{MMIO} in \ac{WASM}.
Therefore, we modified the store and load operations within the \ac{WASM} runtime.

The \ac{WASM} runtime performs a bounds check when executing a store or load instruction.
Since \ac{IO} register addresses (casted to unsigned integers) are located beyond the linear memory, this check will inevitably result in a fault.
If the bounds check fails, \system determines whether the address corresponds to an exposed register address and, if it does, reauthorizes the respective instruction.
This approach is illustrated on the right of \prettyref{fig:approaches}.
Similar to the \ac{RAPI} approach, the \ac{MMIO} method also uses masks to restrict access to specific bits within a register.
When operating in \userspace, \system utilizes a system call to read from or write to the register address.

\paragraph{Hardware acceleration}

\newcommand{\contextswitchbrace}[1]{%
    \scalebox{1}{#1}%
}
\newcommand{\contextswitches}{%
    \mbox{%
        \contextswitchbrace{$[$}%
       user\,%
        $\rightleftarrows$%
        \contextswitchbrace{$]$}%
    }\,%
    \ac{WASM}\,%
    $\rightleftarrows$\,%
    kernel%
}

\begin{wrapfigure}{r}{.5\textwidth}
    \vspace{-0.7cm}
   \includegraphics[width=\linewidth]{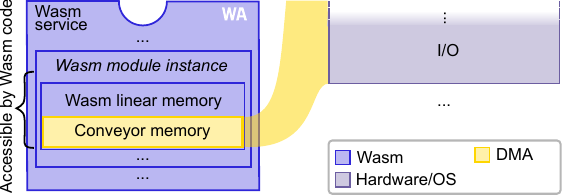}
   \caption{
        The conveyor memory precedes the \ac{WASM} linear memory allowing synchronization with the \acg{OS} \ac{MMIO} leveraging the \ac{DMA} controller.
    }
   \label{fig:memory-setup-dma}
\end{wrapfigure}

As demonstrated in our evaluation in \prettyref{sec:eval}, the primary cause of increased \ac{IO} latency in any solution is context switches {(\contextswitches)}.
To address this issue, we utilize the \ac{DMA} controller.
Additionally, we introduce the concept of conveyor memory, as shown in \prettyref{fig:memory-setup-dma}, which virtually precedes the linear memory of the \ac{WASM} module.
We modify the default \ac{WASM} bounds check to test against a shifted boundary encompassing both the \ac{WASM} linear memory and the conveyor memory.

When resolving register labels to their physical addresses, \system writes these addresses into the \ac{DMA} configuration instead of replacing the dummy registers.
In contrast, the dummy registers are mapped to addresses within the conveyor memory, ensuring that any memory access made by the \service remains within its accessible memory area.
The indirect \ac{DMA} configuration effectively mitigates \ac{DMA} attacks.

During \ac{WASM} execution, the \ac{DMA} controller continuously polls the registers and conveyor memory, promptly updating any changes.
Operating in privileged mode, the \ac{DMA} controller can access both memory regions, \ac{MMIO} in the kernel and \ac{WASM}, potentially in user space.
Furthermore, the inherent overhead of the \ac{WASM} interpretation with multiple cycles per \ac{WASM} instruction hides memory latencies.

\subsection{Interrupt handling}

\system handles interrupts by directly writing the \system prologue to the \ac{IVT}.
Upon the arrival of an interrupt, the \system prologue sequences the \acp{IRQ} by adding them to the epilogue queue.
The highest-priority \ac{OS} thread waits for items in this queue.
Next, the \system epilogue buffers copy data into the interrupt data buffer and invokes the \ac{WASM} epilogues in order of their priority.

When registering an interrupt, the \service signals the index of the corresponding function in the \ac{WASM} internal function table via the \system provided \ac{API}.
Additionally, the data source from which to copy must be defined when employing the data copying mechanism to minimize latency.
The target location in linear memory must also be specified for the \ac{RAPI}, along with the data source.
The \system epilogue then directly copies the data into this designated location.
For \ac{MMIO}, the data is copied into a buffer.
This buffer is located in user space if \system is in user space.

Whenever the \ac{WASM} epilogues are invoked, a new execution environment is created for each \ac{WASM} module instance.
Therefore, the context of the current \service is preserved while allowing global data to be shared through the \ac{WASM} module instance, including global variables and linear memory.
As a result, the executing thread operates outside of \ac{WASM}; the threads proposal~\cite{wattWebAssemblyProposalThreads2024} is not used.

\section{Evaluation}\label{sec:eval}

To assess the efficiency of \system, we aim to demonstrate that \ac{WASM}-based isolation with \system, when executed within the kernel, exhibits overheads and \acp{WCET} comparable to traditional approaches.
We show that upper execution time bounds may be found that enable predictable initial firmware design.
Additionally, we illustrate that overheads can only be managed with hardware support when using untrusted runtimes.

We evaluate by \begin{enumalpha}
    \item externally measuring the worst-observed execution times and
    \item recording the corresponding instruction paths.
\end{enumalpha}
By combining these recorded paths with the source and machine code, we derive the necessary (worst-case) path component for \ac{WCET} analysis, which we denote as \emph{instruction counts}.
When combined with a comprehensive cost model, these instruction counts can effectively be utilized to determine the \acp{WCET}~\cite{liPerformanceAnalysisEmbedded1995}.
However, since the cost model must incorporate a representation of the runtime, this aspect falls outside the scope of our current work.
Instead, our focus is on analyzing loop bounds and, in conjunction with the timing measurements, estimating the real-time behavior and its scaling based on these bounds.

We evaluate two scenarios.
In the first scenario, we assume that the interpreter is trusted, for example, through static verification (see \prettyref{sec:isolation_in_wasm}).
In this case, we compare the execution of \services, interpreted by the \system runtime operating within the kernel, to traditional isolation techniques.
Specifically, we compare with address space separation using Zephyr's \userspace, which employs the \ac{MPU} as a hardware protection mechanism.

In the second scenario, where we do not trust the runtime, it becomes evident that executing programs in an environment isolated by both \ac{WASM} and \userspace leads to suboptimal performance compared to executing solely within \userspace.
Therefore, we compare execution in \userspace, specifically evaluating the impact of \ac{DMA} acceleration both with and without its utilization.

We conduct our evaluation on an {STM32WL55JC} evaluation board that features a Cortex-M4 \ac{CPU} running at \unit{48}{MHz}.
The board is equipped with \unit{64}{KiB} of \ac{SRAM}, and the peripheral buses also operate at \unit{48}{MHz}.
To avoid any potential probe effects, we use a Saleae Logic Pro~16 logic analyzer, which operates at a sampling rate of \unit{250}{MHz}, for external measurements instead of relying on software instrumentation.

\subsection{Synchronous Peripheral Interaction}\label{sec:eval_synchronous_peripheral_interaction}

We measure the roundtrip time required to set and reset a \ac{GPIO} pin (the \emph{measurement pin}) to analyze the costs associated with synchronous peripheral interactions.
\prettyref{fig:measurement} illustrates the measurement process.
The \service executes the software instructions at the \ac{WASM} level to change the pin state, \eg using the \ac{RAPI} call or writing to a memory location.
Consequently, the \ac{WASM} context is exited, dependent on the selected approach.
\system verifies the memory access and, if successful, transitions into the \ac{OS} (kernel).
If \system's runtime is untrusted and \system operates in user space, transitioning happens through a system call.
Following this, the manipulation of the register occurs first at the software level and then at the hardware level.
Afterward, each invoked function returns.
This return process is significant, particularly at both the \ac{OS} (in user space) and \ac{WASM} levels, as it involves extensive copying.
The same procedure applies when resetting the \ac{GPIO} pin.

\begin{fakewrapfig}
The subsequent measurement focuses on the duration between the first \ac{WASM} \ac{IO} instruction and the externally observable event, namely, the signal edge.
Therefore, the actual program in \prettyref{fig:measurement} is omitted (\texttt{NOOP}).
We only present the \acp{WOET}.
It is important to note that the measured instruction counts are not directly equivalent to the time measurements, as shown in \prettyref{fig:measurement}.
While they provide a more detailed categorization of instructions, they lack return paths and hardware latency.

The execution of synchronous \ac{IO} interactions using the Zephyr \ac{API} natively in kernel space yielded a time of \unit{4}{\upmu{}s}.
However, the execution of untrusted code in kernel space contradicts the objective of this work and will, therefore, be excluded from further discussions.

\torightfromhere

\noindent
\includegraphics[width=\linewidth]{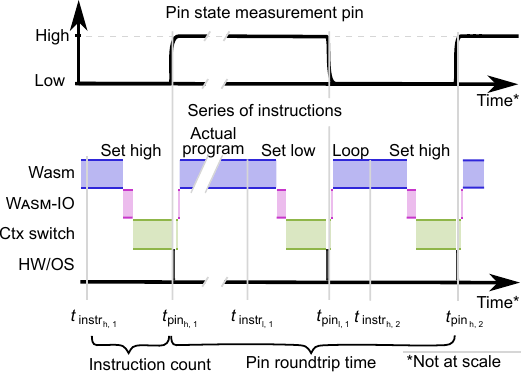}
\captionof{figure}{Schematic of the measurement process for synchronous peripheral interaction leveraging a \ac{GPIO}.}
\label{fig:measurement}

\end{fakewrapfig}

\paragraph*{Trusted Runtime}

\begin{figure}%
    \begin{subfigure}{.49\textwidth}%
        \includegraphics[width=\linewidth]{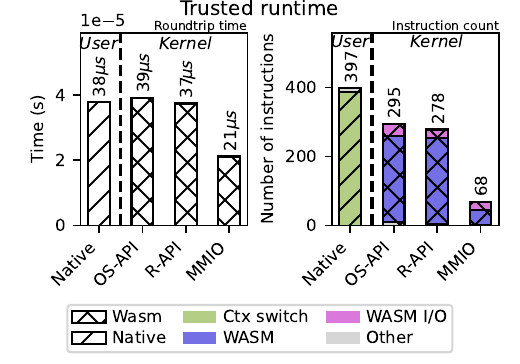}%
        \caption{Measurement \ac{GPIO} pin roundtrip time and instruction count executed in \kernelspace.}%
        \label{fig:pin_roundtrip_time_trusted}%
    \end{subfigure}
    \hfill{}%
    \begin{subfigure}{.49\textwidth}%
        \includegraphics[width=\linewidth]{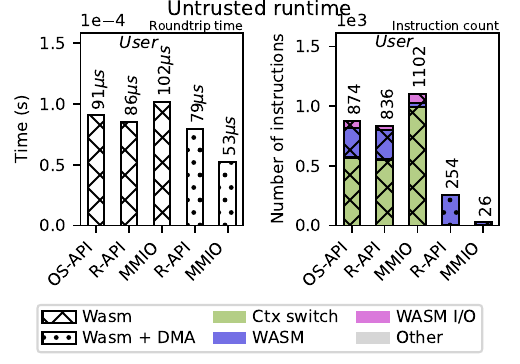}%
        \caption{Measurement \ac{GPIO} pin roundtrip time and instruction count executed in \userspace.}%
        \label{fig:pin_roundtrip_time_untrusted}%
    \end{subfigure}%
\end{figure}

\prettyref{fig:pin_roundtrip_time_trusted} presents the findings of a trusted runtime, whereby the \services are executed in \kernelspace.
We measured the roundtrip time for all three presented approaches: the \ac{OSAPI}, the \ac{RAPI}, and the \ac{MMIO}.
The \ZAPI leveraging \userspace isolation, serves as a baseline and is denoted as \emph{Native}.
As illustrated, the \ac{OSAPI} and the \ac{RAPI} are approximately equivalent in duration to the \ZAPI.
The \ac{MMIO} solution is considerably faster than the others, including the baseline.

An analysis of the instruction counts depicted on the right side of \prettyref{fig:pin_roundtrip_time_trusted} (compare with \prettyref{fig:measurement}) indicates that the extended execution time of the native baseline is predominantly due to context switching.
The \ac{OSAPI} and \ac{RAPI}  figures reveal a substantial number of instructions related to the \ac{WASM} runtime, mainly arising from the execution of imported functions.
This process includes the resolution of function targets, the creation of new \ac{WASM} call stack frames, and the conversion of function arguments.
Conversely, the \ac{MMIO} approach involves only the basic \ac{WASM} bounds check, resulting in markedly reduced execution time and a lower instruction count.
\system's contribution stems from the device handle lookup or the address check.
The slight variations between the \ac{OSAPI} and \ac{RAPI} can be ascribed to the \ac{OSAPI}'s additional kernel-level call to the \ZAPI.

The measurement configuration can be attributed to the discrepancies between timing and instruction counts (\ac{OSAPI} and \ac{RAPI} vs. \ZAPI).
As illustrated in \prettyref{fig:measurement}, the timing metrics encompass the return paths of the \ac{IO} instructions.
While these return paths are negligible in the context of the Zephyr implementation, they significantly impact the \ac{WASM} implementations, where the process of freeing call frames, copying return values, and executing loops within \ac{WASM} contributes additional instructions.
Although the timing measurements capture these intricacies, the corresponding instruction counts do not reflect them.

Our analysis of real-time capabilities for synchronous peripheral interactions emphasizes the substantial dependency on the \ac{WASM} runtime.
However, we regard this dependency as a component of the cost model.
An examination of the recorded traces further indicates that the contribution of \system to the worst-case instruction counts is significantly lower than the \ac{WASM} runtime.
Manual code analysis reveals a linear behavior, scaling by the number of exposed registers across all three methodologies due to searching through these registers.
The current implementation is constrained by a configurable maximum limit.

\paragraph*{Untrusted Runtime}

In the second scenario, the \ac{WASM} runtime is not trusted and, therefore, executes in user space.
\prettyref{fig:pin_roundtrip_time_untrusted} depicts the roundtrip times on the left and instruction counts on the right.
As anticipated, the \acp{WOET} are considerably worse than utilizing the \ZAPI directly from user space (\unit{38}{\upmu{}s}).
Additionally, the \ac{MMIO} solution performs slower than all other options due to the extra bookkeeping data that needs to be copied to kernel space%
\footnote{
The \ac{MMIO} approach entails the creation of additional temporary kernel space copies of \ac{WASM} bookkeeping data, as memory offsets must be validated in the system call.
}.

The \ac{DMA}-accelerated approaches show better performance compared to the non-\ac{DMA} versions, yet they do not surpass the \ZAPI in user space.
The primary factor contributing to overhead is the context switch.
The instruction counts for \ac{DMA} acceleration indicate a reduced \ac{CPU} burden, as the \ac{DMA} unit assumes responsibility for copying.

Upper bounds depend on context switch intrinsics, particularly copying portions of the \ac{WASM} structures to the kernel space, which exhibit a constant size%
\footnote{
Preventing \ac{TOCTOU} attacks requires copying the module instance and execution environment structures and the memory structure for the \ac{MMIO} approach.
}.
The contribution of \system remains unchanged.
For the \ac{DMA} approach, the cost model must account for the external controller, including bus communication.

\subsection{Interrupt handling}

\begin{wrapfigure}{R}{.5\textwidth}
    \includegraphics[width=\linewidth]{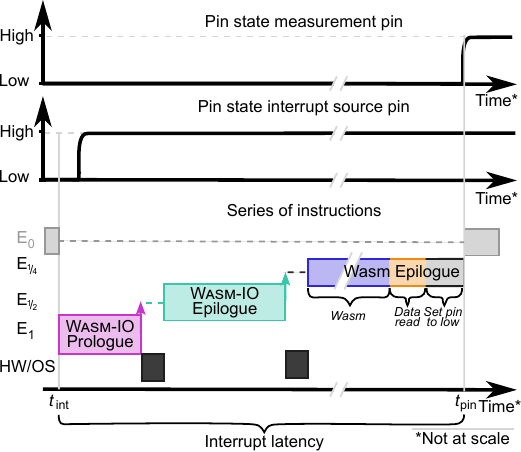}
    \caption{Schematic of the measurement process for interrupts.}%
    \label{fig:measurement-interrupt}
\end{wrapfigure}

\prettyref{fig:measurement-interrupt} illustrates the measurement process used to determine interrupt latencies.
The measurement starts with the rising edge at the interrupt source pin triggering an interrupt.
This spawns the \system prologue.
Following this, the \system and \ac{WASM} epilogues execute, reading the provided data and activating the measurement pin to signal the data reception.
The Zephyr baseline performs the same procedure, utilizing a native callback function to activate the pin after the data is loaded.
To calculate the interrupt latency, we take the difference between the trigger edge and the receive edge and subtract the previously measured pin roundtrip time.
The gray boxes at the bottom of \prettyref{fig:measurement-interrupt} illustrate the limitations of the instruction count measurement procedure.
They indicate the instructions for rescheduling and thread creation that are not included in the instruction count results.
Besides, the \ac{OSAPI} does not provide functions for arbitrary data access, which is why it is excluded from the measurement.

\paragraph*{Trusted Runtime}

\begin{figure}[t]%
    \begin{subfigure}{.49\textwidth}
        \includegraphics[width=\linewidth]{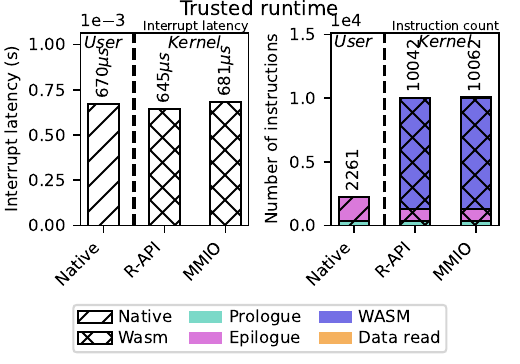}%
        \caption{Interrupt latency and instruction counts with the \ac{WASM} runtime executed in \kernelspace.}%
        \label{fig:interrupt_latency_trusted}%
    \end{subfigure}%
    \hfill{}%
    \begin{subfigure}{.49\textwidth}
        \includegraphics[width=\linewidth]{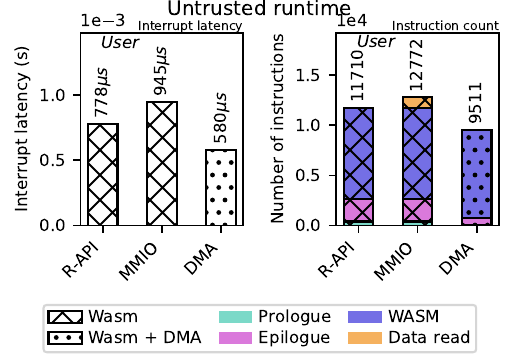}%
        \caption{Interrupt latency and instruction counts with the \ac{WASM} runtime executed in \userspace.}%
        \label{fig:interrupt_latency_untrusted}%
    \end{subfigure}%
\end{figure}

\prettyref{fig:interrupt_latency_trusted} presents the findings from the trusted runtime.
The left graph shows that all interrupt latencies, including the baseline, are comparable.

On the right side, the instruction counts indicate that the duration of the prologue is consistent across all approaches and is relatively brief compared to other components.
The unresponsive time spent in the prologue is less than \unit{25}{\upmu{}s}, while the epilogue for the Wasm approaches takes less than \unit{63}{\upmu{}s}%
\footnote{
These figures do not account for two kernel thread switches (see \prettyref{fig:measurement-interrupt}), as their instruction counts are not measurable in our setup.
Including these would further lower the results.
We only used our \ac{WASM} measurements for calculations, as the absence of context switches invalidates the instruction count measurements for the Zephyr approach.
}.
In all cases, most instructions are related to the transition into the protected domain.
The \system epilogue for the Zephyr baseline takes longer because an additional user thread needs to be initialized to execute the registered callback.

The upper latency bounds are again dependent on the \ac{OS}, its context switch and scheduling implementation, and the invocation of the \ac{WASM} runtime.
We consider these to be research on its own.
The contribution of \system for each \service is directly proportional to the number of registered \ac{WASM} epilogues and the number of data items registered for copying into each linear memory.
There are configurable maxima that constrain both.

\paragraph*{Untrusted Runtime}

It is evident that in the case of an untrusted runtime, the observed latencies are higher, as confirmed by \prettyref{fig:interrupt_latency_untrusted}.
Despite the fact that the \ac{DMA}-accelerated approach is faster than the presented ones for the trusted runtime, real-time bounds once more hinge on the external hardware module, which requires a detailed cost model.
It should be noted that neither the \system prologue nor the \system epilogue is affected by the runtime trustworthiness.

\subsection{Case Study: \acs{SPI}}\label{sec:case-study-spi}

\begin{wrapfigure}{R}{.5\textwidth}
    \includegraphics[width=\linewidth]{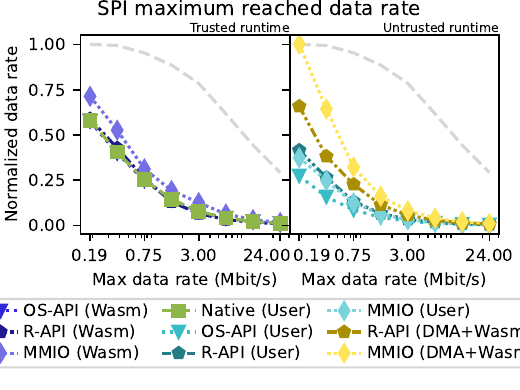}%
    \caption{Reached \ac{SPI} data rate over configured data rate.}%
    \label{fig:spi_maximum_data_rate}%
\end{wrapfigure}

To evaluate the practicality of \system, we developed an exemplary hardware driver in \ac{WASM} that implements an \ac{SPI} bus driver.
\ac{SPI} is a widely accepted standard, and it has a well-established reference implementation within the \ac{OS} and available decoders for verification and measurement purposes.
However, it is important to note that the basic functionality, like an \ac{SPI} driver, is typically embedded in the initial firmware and is, therefore, \emph{not} implemented as a \service.
Our use of this driver is solely for evaluation purposes.

We utilize \unit{1}{KiB} of data organized into \unit{16}{bit} words.
We configure varying data rates ranging from \unit{18.75}{kHz} to \unit{24}{MHz}, equivalent to half of the \ac{CPU} clock frequency.
For measurement, we use the process presented in \prettyref{fig:measurement}.
We set the measurement pin to high before starting the \ac{SPI} transaction and reset it when it is complete;
\ie the \emph{actual program} of \prettyref{fig:measurement} represents the \ac{SPI} transaction.
We subtract the \ac{GPIO} roundtrip time afterward.

\prettyref{fig:spi_maximum_data_rate} presents the results.
The corresponding measured data rate is shown on the y-axis for each configured data rate depicted on the x-axis.
The values have been normalized with respect to the configured rate.
The gray dashed line devoid of markers depicts the execution with the native \ZAPI in kernel space, indicating the attainable maximum data rate.
Even this most performant implementation suffers from a reduction in data rate to 99.5\% already at \unit{37.5}{kHz}, the second lowest data rate.
This reduction can be attributed to the relatively low \ac{CPU} clock frequency compared to the high \ac{SPI} clock frequency.

The values of the trusted runtime are consistent with those of the measurement in \prettyref{sec:eval_synchronous_peripheral_interaction}.
As expected, the \ac{MMIO} approach performs best.
Notably, it also performs better than the \ac{OSAPI}.
Nevertheless, none of these approaches reaches more than 71\% of the possible data rate, even if the lowest data rate is configured.
Interestingly, in the untrusted case, the \ac{OSAPI} performs worse than both the \ac{RAPI} and \ac{MMIO}, which contradicts our previous measurement.
This discrepancy is due to different implementations of \ac{SPI} \ac{OSAPI} and \ac{GPIO} \ac{OSAPI}.
The former copies references to the \ac{WASM} memory, which requires an additional copy of the \ac{WASM} linear memory.
The latter copies the value of an integer (high or low) to the \ac{OS} via the stack of \ac{OS} thread, which is significantly faster.

The \ac{DMA} solutions show remarkably better results as they can copy \ac{SPI} registers without involving the \ac{CPU}, thus avoiding transitions from the protection domain for each byte.
This allows the full data rate, at least for \unit{18.75}{kHz}z, for the \ac{MMIO} approach.

The overall results show that instrumentation of \ac{IO} is feasible, but it comes with a considerable, yet bound overhead.
The \ac{MMIO} approach outperforms the traditional user space isolation in synchronous \ac{IO} and in our \ac{SPI} case study but requires trusted runtimes.
Asynchronous \ac{IO} shows comparable overheads over all solutions in the trusted case.
The untrusted case reveals significant performance impacts which impede usage in praxis.
\ac{DMA} acceleration decreases the overall overheads.
However, it is accompanied by the need for an increasingly complex hardware cost model.

\section{Discussion and Usage}\label{sec:discussion}

This section discusses the impact, practicality, and application of \system for designing \ac{WASM}-based systems in the automation industry, particularly enabling the addition of peripherals and drivers post-deployment.
We envision offering access to unused pins and connectors through modular plugs integrated into the casing, designed with industrial-grade safety measures to mitigate risks associated with overvoltage and other physical threats.
Functional extensions can be loaded via the \services that were presented.
Services will interact with the firmware through the \ac{OSAPI} to enable typical sensing tasks, contingent upon temporal isolation not being compromised.
Examples include packet-based communication, where protocols ensure transmission completeness, allowing the firmware to intercept \service execution as needed.
Replacing a firmware driver with a \service driver is intentionally unsupported.

Choosing a proper scheduling policy depends on the actual use case.
Systems operate either in an event-driven manner using interrupts, \eg for emergency shutdowns, or periodically, \eg in a sense/listen-and-act loop with preemptive static scheduling.
\system ensures temporal isolation in both types of systems.
For event-driven systems, the \system epilogue executes with a lower priority than firmware epilogues.
For periodic systems, a periodic server processes queued \system epilogues, also with lower priority than critical firmware epilogues.
To allow multiple \ac{WASM} application services to utilize the functionality provided by a peripheral, \services may be designed to resemble microservices.
A dedicated proxy \ac{WASM} driver service solely manages this peripheral, maintaining the exclusive device mapping, and mitigates indeterminism raised by concurrency issues.
Currently, \system does not facilitate communication between \services.
However, providing a host \ac{API} for inter-\service communication is feasible, even with timing bounds if required.

Overall, \system establishes effective conventions and guarantees for future-proof automation systems.
It enables a clear separation of responsibilities across different dimensions by dividing the development process into two phases:
firmware development and the extension period.
The responsibility for firmware development is enforced but also limited to the \ac{OEM}, while the extension period is solely assigned to the operators.
\system maintains this separation at the device level by utilizing memory isolation through \ac{WASM} and timing bounds for usable interrupts, which enables temporal isolation through a proficient scheduling policy.
Furthermore, the obligation to preserve records may be delegated from \acp{OEM} to operators after legally mandated periods, thereby mitigating conflicts of interest that arise from being the single entity facilitating undertain follow-up modifications and maintaining cost-effectiveness.
In contrast, operators retain the autonomy to choose whether and how to enhance their manufacturing systems, including the option to engage vendors other than the initial \ac{OEM}.

Our evaluation indicates that, although \system is compatible with alternative isolation approaches, it incurs notable overheads.
However, the computational overheads of \ac{IO} operations are not the predominant aspects of program latencies.
The \ac{IO} speed of sensors and actuators typically depends more on the duration of the physical processes—such as collecting temperature data or operating motors—rather than communication speed.
Additionally, our target industry, the process industry, is characterized by inherently slow processes that do not necessitate exceptionally rapid \ac{IO} speeds.
Consequently, this aspect was not the primary focus of our work.
In practice, only a limited number of registers (two to four) are utilized after initialization.
As they are sorted by access frequency, experienced overheads are lower than the presented worst-observed execution times.

We conclude that \system effectively addresses the challenges of arbitrary peripheral \ac{IO} maintaining deterministic system behavior, as discussed in \prettyref{sec:introduction}.
It offers practical solutions for the process industry and enables post hoc driver extensions, extending the lifespan of long-life devices even further.
Additionally, it brings the advantages of container deployments known from cloud environments one step closer to industrial shop floors.

\section{Related Work}\label{sec:related_work}

The topic of enabling device \ac{IO} in virtualized systems has been subject of extensive research for over a decade.
This is particularly relevant for hypervisors, where \system can be compared with, when coinciding \services as \acp{VM}.
The existing body of work can be classified into two main categories: pass-through \ac{IO} and paravirtualized \ac{IO}.
In the former, the virtual machine is granted exclusive access to the device.
Tu~\etal~\cite{tuComprehensiveImplementationEvaluation2015} present a KVM-based implementation that exposes a virtual interrupt controller register to the \ac{VM}, ultimately enabling preemption of \ac{VM} execution.
In the XEN-based approach proposed by Yoo~\etal~\cite{yooStepSupportRealTime2009}, the hypervisor is responsible for preempting \ac{VM} execution.
Paravirtual \ac{IO} employs a driver architecture comprising a backend and a frontend driver.
In XEN, proposed by Barham~\etal~\cite{barhamXenArtVirtualization2003}, interrupts are stored and can be collected and processed per \ac{VM} whenever the \ac{VM} is scheduled again.
Chen and Wang~\cite{chengOffloadingInterruptLoad2016} further extend this approach by transferring interrupt handling to the hypervisor, which schedules the workload on free virtual \acp{CPU}.
Ahmad, Gulati, and Mashizadeh~\cite{ahmadVICInterruptCoalescing2011} implemented an interrupt handling mechanism for VMware~ESX that is analogous to that used by XEN.
This utilizes a shared queue where \acp{VM} could fetch their interrupts.
In their BlueIO, Jiang~\etal~\cite{jiangBlueIOScalableRealTime2019} employ the use of \acp{FPGA} to facilitate \ac{IO} interaction.

In the context of software virtualization, where \ac{WASM} belongs to, peripherals are typically utilized by exposing the \ac{IO} \ac{API} provided by the \ac{OS}~\cite{singhWARDuinoDynamicWebAssembly2019,liBringingWebassemblyResourceconstrained2022} or by adding and exposing more high-level interaction, also managed by the \ac{OS}~\cite{liuAerogelLightweightAccess2021,liWiProgWebAssemblybasedApproach2021}.
In the case of \ac{AoT} compilation, native functions can be replaced directly with assembly instructions, thereby enabling the utilization of \ac{IO}~\cite{liBringingWebassemblyResourceconstrained2022}.
These limitations restrict the interactions that can be performed to those of the \ac{OS}.
Moreover, any isolation guarantees provided by \ac{WASM} are either lost or deferred to the underlying \ac{OS}.
\ac{WASI}~\cite{gohmanWebAssemblyWASIV0222024}, used in other approaches~\cite{wenWasmachineBringIoT2020}, provides standardization efforts to establish a uniform system interface, although it is currently not specified and not yet tailored to embedded systems~\cite{sigembeddedWASIWasmSystem2025}.

Other software isolation methods for embedded systems employ analogous techniques for \ac{IO}.
The range of solutions available in the context of Java extends from interaction on memory-mapped peripherals~\cite{stilkerichTailormadeJVMsStatically2012} to high-level driver access provided by the \ac{JVM}~\cite{aslamOptimizedJavaBinary2010,brouwersDarjeelingJavaCompatible2008}.
Zandberg~\etal~\cite{zandbergFemtocontainersLightweightVirtualization2022} base their isolation on eBPF, utilizing the interface provided by the OS for potential interaction with the hardware.
Tok~\cite{levyMultiprogramming64kBComputer2017}, an \ac{OS} written in Rust, translates memory mappings into type-safe Rust structs.
Hardware-based isolation leverages Arm's TrustZone for multiple permission levels and partitioned hardware~\cite{pintoVirtualizationTrustZoneEnabledMicrocontrollers2019,jiangMCSIOVRealTimeVirtualization2019}, even supporting interrupts~\cite{panSBIsApplicationAccess2022}.

\section{Conclusion}\label{sec:conclusion}

In this paper, we presented how \system enables software extensions of long-lifecycle constrained devices, including low-level device drivers.
Our approach facilitates the integration of cross-platform hardware requirements by labeling peripheral dependencies. Our prototype derives these labels from immutable \devicetree nodes.
The corresponding \ac{MMIO} regions, which share these designated labels, are rendered available to \services, provided that they are marked accordingly at firmware design time by our novel \devicetree binding.
\system enables synchronous \ac{IO}, implementing, an \ac{OS}-guided approach and two novel approaches, one \acf{MMIO}-based approach and the \acf{RAPI} approach.

Moreover, we introduced support for interrupt handling into \ac{WASM}.
We extended the traditional system model by another priority level E$_{\nicefrac{1}{4}}$, dedicated explicitly to \services.
Thereby, \system provides a means to decouple the timely influence associated with a \service from the remaining system.
With this separation, we introduce
\begin{enumalpha}
    \item a constant overhead per interrupt handler per \service on E$_0$ and E$_{\nicefrac{1}{2}}$ and
    \item limits on the number of handlers and \services to constrain the overall execution time during firmware creation.
\end{enumalpha}
The resulting upper bound is contingent upon the scheduling mechanism used.
One possible solution could be the use of a server to serve the \ac{WASM} epilogues on E$_{\nicefrac{1}{4}}$.

In our evaluation, we presented how hardware interaction and interrupt latencies and the costs associated with containerization can be pushed to practical levels while preserving strong isolation.
The results demonstrate that the \ac{MMIO}-based approach is advantageous when writing to registers.
Our case study demonstrates that the isolation costs are non-negligible yet bound to fixed values.
Hardware acceleration, \eg by the proposed \ac{DMA} approach, helps significantly increase data rates but has the disadvantage of a more complex hardware cost model.

It can be concluded that hardware configuration and instrumentation can be performed from a \service in a bounded time.
However, hardware acceleration should be taken into account for performance reasons when accordable with real-time demands.

\system effectively tackles the initial challenges of post hoc firmware extensions, including integrating drivers for novel peripherals.
Thereby, \system extends the lifespan of long-life devices, enabling future compatibility in automation industry.
Moreover, \system seamlessly brings the conveniences of containerization -- with benefits widely recognized in cloud development -- bringing them one step closer to the shop floor.

\begin{acks}
This work was funded by the German Federal Ministry of    Education and Research (BMBF) under grant number 16ME0454 (EMDRIVE). To assist the writing process, {Grammarly AI} was used.
\end{acks}

\bibliographystyle{ACM-Reference-Format}
\bibliography{paper}

\end{document}